\documentclass[pra,onecolumn,superscriptaddress]{revtex4-1}%
\usepackage[T1]{fontenc}
\usepackage[english]{babel}
\usepackage{graphicx}
\usepackage{hyperref}
\usepackage{amsmath}
\usepackage{amsfonts}
\usepackage{amssymb}
\hypersetup{pdfstartview=FitH,	pdfpagemode=UseNone}

\newcommand{\tauc}{\tau_{\mathrm{corr}}}
\newcommand{\der}[2]{\frac{\mathrm{d}#1}{\mathrm{d}#2}}
\newcommand{\derr}[2]{\frac{\mathrm{d}^2#1}{\mathrm{d}#2^2}}
\newcommand{\rder}[2]{\frac{\partial#1}{\partial#2}}

\begin{document}

\title{Frequency shifts and relaxation rates for spin-1/2 particles moving in electromagnetic fields}
\author{G. Pignol}
\email{pignol@lpsc.in2p3.fr}
\author{M. Guigue}

\affiliation{LPSC, Universit\'e Grenoble-Alpes, CNRS/IN2P3, Grenoble, France}
\author{A. Petukhov}
\affiliation{Institut Laue Langevin, 6 Rue Jules Horowitz, 38000 Grenoble, France}
\author{R. Golub}
\affiliation{Physics Department, North Carolina State University, Raleigh, NC 27965}
\date{\today}

\begin{abstract}
We discuss the behavior of the Larmor frequency shift and the longitudinal relaxation rate due to non-uniform electromagnetic fields  on an assembly of spin 1/2 particles, in adiabatic and nonadiabatic regimes.
We also show some general relations between the various frequency shifts and between the frequency shifts and relaxation rates.
The remarkable feature of all our results is that they are obtained without any specific assumptions on the explicit form of the correlation functions of the fields. 
Hence, we expect that our results are valid for both diffusive and ballistic regimes of motion and for arbitrary cell shapes and surface scattering.
These results can then be applied to a wide variety of realistic systems. 
\end{abstract}
\maketitle

\section{Introduction}

The behavior of a system of spins interacting with static and time varying magnetic fields is a very broad topic and has been the subject of intense study for decades. 
A very important application is to the study of spins interacting with the randomly fluctuating fields associated with a thermal reservoir. 
Bloembergen, Purcell and Pound \cite{Bloembergen1948} have treated this problem using physical arguments based on Fermi's golden rule and showed that the relaxation induced by the fields associated with a thermal reservoir is proportional to the power spectrum of the fluctuating fields evaluated at
the Larmor frequency, $\omega_0=\gamma B_0,$ (where $\gamma$ is the
gyromagnetic ratio and $B_0$ is an applied constant and uniform field),
which is given by the Fourier transform of the auto-correlation function of
these fluctuating fields evaluated at $\omega_0$. Wangsness and Bloch
\cite{Wangsness1953} and then Bloch \cite{Bloch1956} approached the
problem using second order perturbation theory applied to the equation of motion of the density matrix and Redfield, \cite{Redfield1957,Redfield1965}
(see also \cite{Slichter1964}) carried this calculation forward to show
that the relaxation, indeed, depends on the spectrum of the auto-correlation
of the fluctuating fields.

Another source of randomly fluctuating fields is the stochastic motion of
spins (e.g. diffusion) through a region with an inhomogeneous magnetic field.
To study this problem Torrey \cite{Torrey1956} introduced a diffusion term
into the Bloch equation applied to the bulk magnetization of a sample containing many spins (Torrey equation). 
Cates, Schaeffer and Happer \cite{Cates1988} then rewrote the Torrey equation to apply to the density
matrix and solved this equation to second order in the varying fields using an
expansion in the eigenfunctions of the diffusion equation. 
McGregor \cite{McGregor1990} applied the Redfield theory to this problem using
diffusion theory to calculate the auto-correlation function of the fluctuating
fields experienced by spins diffusing through a (uniform gradient) inhomogeneous field. 
Recently Golub \textit{et al.} \cite{Golub2010c} showed that these two approaches \cite{Cates1988,McGregor1990} are identical. 
A useful review of the field is \cite{Nicholas2010}.

Another problem that can be treated by these methods is the case of a gas of
spins contained in a cell subject to inhomogeneous magnetic fields and a
strong electric field as in experiments to search for a non-zero electric
dipole moment (EDM) of neutral particles such as the neutron \cite{Lamoreaux2009} or
various atoms or molecules \cite{Eckel2013}. 
This was shown by Pendlebury \textit{et al.} \cite{Pendlebury2004} using a 
second order perturbation approach to the classical Bloch equation, 
to lead to an unwanted, linear in electric field, frequency shift, 
(often called a 'false EDM' effect) which can be the largest systematic error in such experiments. 
Lamoreaux and Golub \cite{Lamoreaux2005b} showed, using a standard density matrix calculation
(Redfield theory), that the 'false EDM' frequency shift is given, 
to second order, by certain correlation functions of the fields seen by the moving particles.

Barabanov \textit{et. al.} \cite{Barabanov2006} gave analytic expressions for the relevant correlation functions for a gas of particles moving in a cylindrical vessel exposed to a magnetic field with a linear gradient along with an electric field. 
Petukhov \textit{et al.} \cite{Petukhov2010} and Clayton \cite{Clayton2011} showed how to determine the correlation functions for arbitrary geometries and spatial field dependence for cases where the diffusion theory applies, while Swank \textit{et. al.} \cite{Swank2012} showed how to calculate the spectra of the relevant correlation functions for gases in rectangular cells in
magnetic fields of arbitrary position dependence even in those cases where the
diffusion theory does not apply. 
Recently Afach \textit{et al} \cite{Afach2015} measured a frequency shift that is linearly proportional to an applied electric field (false electric dipole moment) for a system consisting of Hg atoms moving in a confined gas exposed to parallel electric and magnetic fields.

Pignol and Roccia \cite{Pignol2012a} have initiated a program to search for universal expressions giving general results valid for all geometries and scattering conditions in the gas and gave such a result for the false EDM effect valid in the nonadiabatic (low frequency) limit. 
Further steps in this direction were taken by Guigue \textit{et. al.} \cite{Guigue2014} who provided a universal result for frequency shifts induced by inhomogeneous fields in the adiabatic (high frequency) limit and for the relaxation rate ($\Gamma_{1}$) in the case where diffusion theory applies.

In this work we extend the search for universal expressions of frequency shifts and relaxation for both the adiabatic and nonadiabatic (high and low Larmor frequency) limits.

\section{Frequency shifts and relaxation rates from Redfield theory}

We consider the case of a gas of spin-1/2 particles inside a trap with a gyromagnetic ratio $\gamma$ evolving in a slightly inhomogeneous
magnetic field $\vec{B}(\vec{r}) = \vec{B}_0 + \vec{b}(\vec{r})$. 
One can define the holding magnetic field $\vec{B}_0 = B_0\vec{e}_{z}$ and the Larmor precession frequency $\omega_0 = \gamma B_0$. 
The inhomogeneities $\vec{b}$ can be taken to have $\langle \vec{b} \rangle = \vec0$ where $\langle \cdots \rangle$ represents the ensemble average over all particles in the trap.
In addition to this inhomogeneity, the particles can move with a velocity $\vec{v}$ in an electric field $\vec{E}$. 
For simplicity, one can consider that the direction of this electric field is aligned with the holding magnetic field: 
$\vec{E} = E\vec{e}_z$. 
These particles will experience an effective motional magnetic field $\vec{E} \times \vec{v}/c^2$. 
The transverse components of the total magnetic inhomogeneity will then depend on the position and the velocity of the particles in the trap
\begin{align}
B_x  &  = b_x - \frac{E}{c^2} v_y \\
B_y  &  = b_y + \frac{E}{c^2} v_x. 
\label{eq:}
\end{align}

These transverse inhomogeneities induce a shift $\delta\omega$ of the precession frequency and a longitudinal relaxation rate $\Gamma_1$. 
Correct to second order in the perturbation, $b$, the frequency shift $\delta\omega$,
the longitudinal relaxation rate $\Gamma_{1}$ and the transverse relaxation rate $\Gamma_2$, involving the Fourier spectra
of the inhomogeneity correlation functions, are given by the Redfield theory
\cite{Slichter1964,Redfield1965,McGregor1990,Guigue2014}: 
\begin{align}
\delta\omega &  = \frac{\gamma^2}{2} \left\{ \mathrm{Re}\left[  S_{xy}(\omega_0) - S_{yx}(\omega_0) \right]  + \mathrm{Im}\left[ S_{xx}(\omega_0) + S_{yy}(\omega_0)\right]  \right\},
\label{eq:dw-Redfielc}\\
\Gamma_1  &  = \gamma^2\left\{  \mathrm{Re} \left[ S_{xx}(\omega_0) + S_{yy}(\omega_0)\right]  + \mathrm{Im}\left[ S_{yx}(\omega_0) - S_{xy}(\omega_0)  \right]  \right\} \\
\Gamma_2  &  = \frac{\Gamma_1}{2} + \gamma^2 S_{zz}(\omega=0)
\end{align}
with
\begin{equation}
S_{ij}(\omega)=\int_0^\infty e^{i\omega\tau}\langle B_i(0)B_j(\tau)\rangle\mathrm{d}\tau. 
\label{eq:1}
\end{equation}
This result is valid in cases where the field fluctuations are stationary in the statistical sense and
where the measurements are made over a time scale $T\gg\tauc$
where the correlation time $\tauc$ is the time scale for which
the correlation functions go to zero. 
In the case of particles evolving in
both an inhomogeneous magnetic field and an electric field, the frequency
shift $\delta\omega$ and relaxation rate $\Gamma_{1}$ can be decomposed as
\begin{align}
\delta\omega &  = \delta\omega_{B^2} + \delta\omega_{E^2} + \delta\omega_{BE},
\label{eq:dw-decomposition} \\
\Gamma_1     &  = \Gamma_{1 (B^2)} + \Gamma_{1 (E^2)} + \Gamma_{1 (BE) ,}
\end{align}
with
\begin{eqnarray}
\delta \omega_{B^2}  &  = \frac{\gamma^2}{2}  & \mathrm{Im} \int_0^\infty e^{i\omega_0\tau} \langle b_x(0)b_x(\tau)+b_y(0)b_y(\tau)\rangle \mathrm{d}\tau,
\label{eq:dw-B2} \\
\delta\omega_{E^2}   &  = \frac{\gamma^2 E^2}{2c^{4}} & \mathrm{Im} \int_0^\infty e^{i\omega_0\tau} \langle v_x(0)v_x(\tau) + v_y(0)v_y(\tau) \rangle \mathrm{d}\tau,
\label{eq:dw-E2} \\
\delta\omega_{BE}  &  = \frac{\gamma^2E}{c^2} & \mathrm{Re} \int_0^{\infty} e^{i\omega_0\tau} \langle b_x(0)v_x(\tau) + b_y(0)v_y(\tau) \rangle \mathrm{d}\tau,
\label{eq:dw-BE} \\
\Gamma_{1 (B^2)}  &  = \gamma^2  & \mathrm{Re} \int_0^\infty e^{i\omega_0\tau} \langle b_x(0)b_x(\tau)+b_y(0)b_y(\tau) \rangle \mathrm{d}\tau,
\label{eq:rel-B2} \\
\Gamma_{1 (E^2)}  &  = \frac{\gamma^2 E^2}{c^{4}} & \mathrm{Re} \int_0^\infty e^{i\omega_0\tau} \langle v_x(0)v_x(\tau) + v_y(0)v_y(\tau) \rangle \mathrm{d}\tau,
\label{eq:rel-E2} \\
\Gamma_{1 (BE)}  &  = \frac{2 \gamma^2E}{c^2} & \mathrm{Im} \int_0^\infty e^{i\omega_0\tau} \langle b_x(0)v_x(\tau) + b_y(0)v_y(\tau) \rangle \mathrm{d}\tau. 
\label{eq:rel-BE}
\end{eqnarray}
These Larmor frequency shifts and relaxation rates cannot be further simplified to a form valid for all values of holding magnetic field and independent of the particle motion in the trap. 
However, due to the properties of the Fourier transform, there are universal relations that hold for all types of particle motion and all shapes of trap geometry for values of Larmor frequency (magnetic field) large and small relative to the inverse transit time of particles across the cell, $\lambda/v$ (see \ below).

\section{Spin dynamics and particle motion regimes}

In general two length scales describe the motion of a gas of particles in a cell: 
(i) the mean free path between particle collisions noted $l_c$, and 
(ii) the mean distance between two points on the wall which can be evaluated by the Clausius expression $\lambda = 4 V/S$ where $V$ and $S$ are the volume and the surface of the cell. 
We define Knudsen's number as $Kn=\frac{l_c}{\lambda}$. 
At high pressure, $Kn \ll 1$: this is the \textit{diffusive regime} where the propagation of the particles is described by the diffusion equation, characterized by the diffusion coefficient $D$. 
At low pressure, $Kn \gg 1$: this is the \textit{ballistic regime} where the particles travel in straight lines across the cell in free molecular flow. 

The correlation time $\tauc$ corresponds to the typical time necessary for a particle to probe the magnetic inhomogeneity. 
Since one usually has to deal with large scale inhomogeneities, $\tauc$ is of the order of the average time between successive collisions with the trap walls. 
Therefore, it depends on the geometry of the trap and on the properties of the particle motion inside this trap. 
In the case of a gas at atmospheric pressure, the correlation time is about $1~\mathrm{s}$ for a cubic trap with $10~\mathrm{cm}$ sides. 
For rarefied gas confined in the trap, $\tauc$ is approximately equal to $1~\mathrm{ms}$. 
This time scale can be compared with the Larmor precession frequency $\omega_0$. 
The limit when $\omega_0$ is much bigger than $1/\tauc$, is called \textit{adiabatic regime}. 
This regime can be interpreted as the particle spins following the local magnetic field. 
It is also valid when the particles are moving slowly in the trap or if they encounter a great number of collisions with other particles between two collisions with walls. 
In contrast, the regime is called \textit{nonadiabatic} if $\omega_0 \tau_c \ll 1$. 
This limit physically appears when the particles are able to probe the whole magnetic inhomogeneity within times shorter than a Larmor period. 
It is also sometimes refered to as the regime of \textit{motional narrowing}. 
This phenomenon can be observed in systems immersed in very weak magnetic fields or if the thermal particles are in a ballistic regime in a small container.

\begin{figure}
\centering\includegraphics[width=0.50\textwidth]{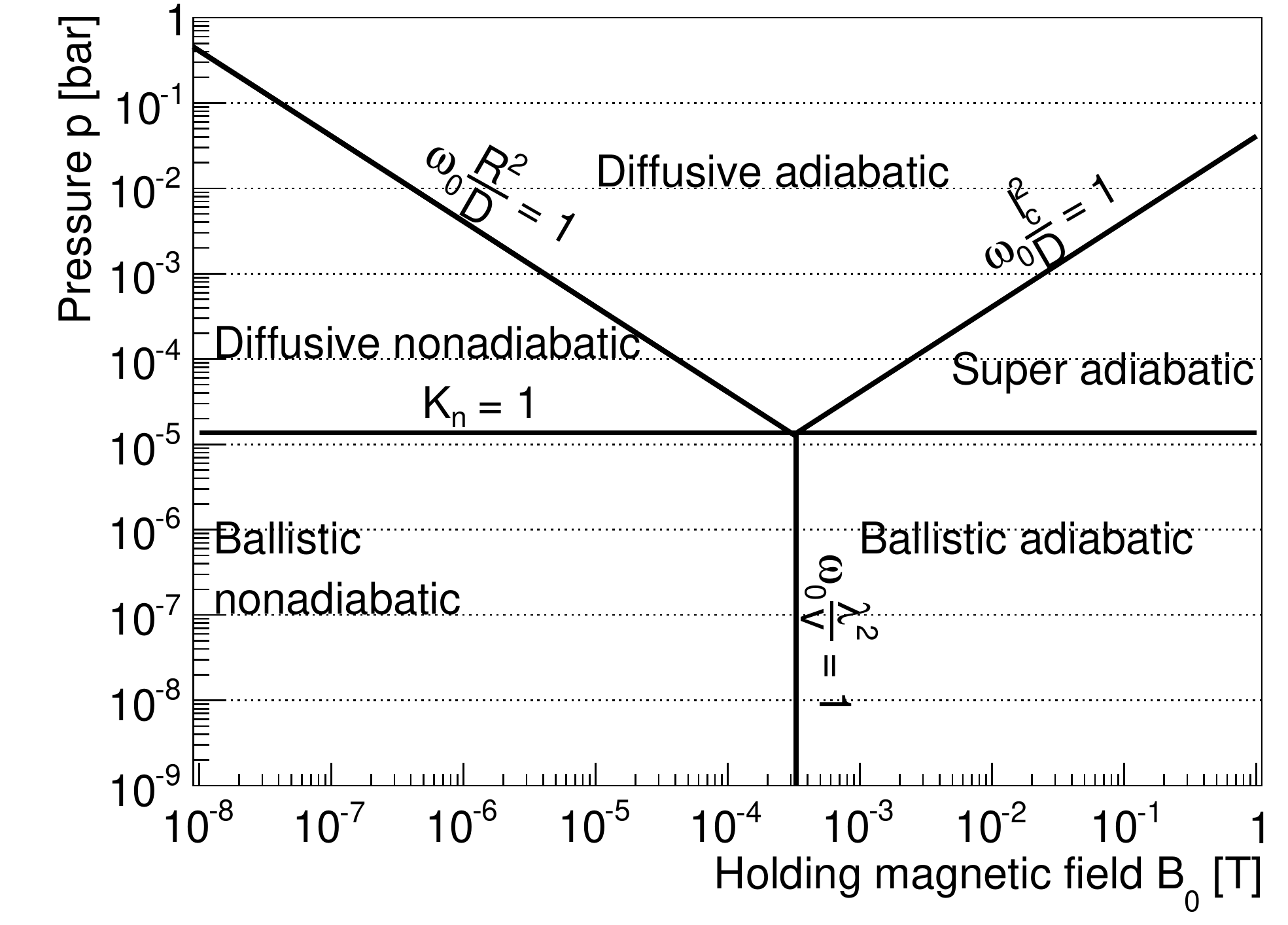}
\caption{Classification of the different regimes for a $R=5$~cm radius spherical cell filled with polarized $^3$He gas as a function of pressure and holding magnetic field.}
\label{fig:classification-regime}%
\end{figure}

For a given trap geometry, these regimes depend on the pressure of the spin gas and on the holding magnetic field. 
Fig. \ref{fig:classification-regime} shows this classification as a function of pressure and holding field for a $^3$He gas contained in a spherical cell with $5~\mathrm{cm}$ radius. 
The \textit{super-adiabatic} regime corresponds to the situation where the gas is in a diffusive regime and the spin motion is adiabatic between two interparticles collisions. 
In this case, we have the condition $\omega _0\tau _{\rm{coll}}\gg 1$, with $\tau_{\rm{coll}}$ the time between two interparticles collisions.
The correlation functions calculated in \cite{Swank2012} are valid in this region as is Eq. (12) in \cite{McGregor1990}.

To illustrate this classification, let us consider some realistic systems of particular relevance for our study. 
The cylindrical RAL/Sussex/ILL trap \cite{Pendlebury2004}, used to measure the neutron EDM, 
contains ultracold neutrons (UCN) and a mercury comagnetometer,
immersed in a $1~\mathrm{\mu T}$ holding magnetic field. 
The small particles number of each species and the size of the trap ($47~\mathrm{cm}$ diameter and
$12~\mathrm{cm}$ height) lead to a large Knudsen's number and so to the
ballistic regime. 
However, the speeds of the two particle species are very different: while the mercury atoms are at thermal equilibrium and have an
average speed of several hundred meters per second, the UCN are moving at a
few meters per second. Therefore the mercury comagnetometer is in the
ballistic nonadiabatic regime whereas UCN are in the ballistic adiabatic limit. 
In the case of a gas at atmospheric pressure, such as a polarized $^3$He gas \cite{Petukhov2010}, in a several $\mu\mathrm{{T}}$ holding field, 
the particle motion follows the diffusion equation and the number of Larmor precessions done by the spins between two collisions with the walls is very high. 
This kind of systems is thus in diffusive adiabatic regime.

As shown in \cite{Pignol2012a,Guigue2014}, the leading order of frequency shifts (\ref{eq:dw-B2}), (\ref{eq:dw-E2}) and (\ref{eq:dw-BE}) for
adiabatic and nonadiabatic regimes can be expressed as powers of $\omega_0 \tauc$ or $1/\omega_0 \tauc$. 
To do so, we apply a succession of integrations by parts of the integrals defining the frequency shifts. 
This is the purpose of the two next sections. 
The first one presents the simplified expression of the frequency shift in the adiabatic regime. 
The nonadiabatic regime will be considered in the second next section. 
The well-known case of uniform magnetic gradients will be discussed in Section VI.

\section{Adiabatic regime: high magnetic field or slow particles, arbitrary fields}

The adiabatic regime corresponds to systems which satisfy $\omega_0\gg 1/ \tauc$.
We want to expand the frequency shift (\ref{eq:dw-decomposition}) in power series of $1/\omega_0$. 
One way to obtain such an expression consists in applying several integrations by parts
\begin{equation}
\int_0^\infty \sin(\omega_0 \tau) f(\tau) \mathrm{d}\tau = 
\left[ \frac{-\cos(\omega_0\tau)}{\omega_0}f(\tau) \right]_0^\infty 
+ \frac{1}{\omega_0}\int_0^\infty \cos(\omega_0\tau) \der{f}{\tau}(\tau)\mathrm{d}\tau, 
\label{eq:IPPcos}
\end{equation}
\begin{equation}
\int_0^{\infty}\cos(\omega_0\tau)f(\tau) \mathrm{d}\tau = \left[  \frac
{\sin(\omega_0\tau)}{\omega_0}f(\tau)\right]  _0^\infty 
- \frac{1}{\omega_0}\int_0^{\infty}\sin(\omega_0\tau) \der{f}{\tau}(\tau)\mathrm{d}\tau. 
\label{eq:IPPsin}
\end{equation}
Equation \eqref{eq:IPPcos} and \eqref{eq:IPPsin} assume that the function $f$ and
its derivative are integrable. 
We denote $\dot{f}=\der{f}{\tau}$. 
Using the fact that the correlation functions go to zero at infinite time, we can write
\begin{eqnarray}
\delta \omega_{B^2} & = & \frac{\gamma^2}{2\omega_0} \langle b_x^2+b_y^2 \rangle 
- \frac{\gamma^2}{2\omega_0^3} \langle b_x(0)\ddot{b}_x(0)+b_y(0)\ddot{b}_y(0) \rangle \nonumber \\
& - & \frac{\gamma^2}{2\omega_0^{3}} \int_0^\infty \cos(\omega_0\tau)\langle b_x(0)\dddot{b}_x(\tau)+b_y(0)\dddot{b}_y(\tau)\rangle \mathrm{d}\tau, 
\label{eq:dw-B2-int}
\end{eqnarray}
\begin{eqnarray}
\delta\omega_{E^2} & = & \frac{\gamma^2 E^2}{2c^{4}\omega_0}\langle v_x^2 + v_y^2 \rangle \nonumber \\
& + & \frac{\gamma^2 E^2}{2c^{4}\omega_0} \int_0^\infty \cos(\omega_0\tau) \langle v_x(0)\dot{v}_x(\tau)+v_y(0)\dot{v}_y(\tau) \rangle \mathrm{d}\tau, 
\label{eq:dw-E2-int}
\end{eqnarray}
\begin{eqnarray}
\delta \omega_{BE} & = & \frac{\gamma^2E}{c^2\omega_0^2}\langle b_x(0)\dot{v}_x(0) + b_y(0)\dot{v}_y(0) \rangle \nonumber \\
& - & \frac{\gamma^2E}{c\omega_0^2}\int_0^\infty\cos(\omega_0\tau) \langle b_x(0)\ddot{v}_x(0)+b_y(0)\ddot{v}_y(0)\rangle\mathrm{d}\tau. 
\label{eq:dw-BE-int}
\end{eqnarray}

Making the reasonable assumption that the correlation functions and their derivatives are continuously decaying to 0 for $\tau\rightarrow\infty$, we apply the Riemann-Lebesgue  lemma \cite{Appel} and arrive at the conclusion that the last terms in Eq. \eqref{eq:dw-B2-int}, \eqref{eq:dw-E2-int}, \eqref{eq:dw-BE-int} go to zero faster than the other terms in each equation.
The frequency shift expressions can be written as (see also Eq. \eqref{eq:Sij-dev}, \eqref{bg1}, \eqref{bg2}):
\begin{equation}
\delta \omega_{B^2} = \frac{\gamma^2}{2\omega_0} \langle b_x^2 + b_y^2 \rangle - \frac{\gamma^2}{2\omega_0^{3}}\langle b_x(0)\ddot{b}_x(0) + b_y(0) \ddot{b}_y(0) \rangle + O \left(  1/(\omega_0 \tauc)^5\right), 
\label{eq:dw-B2-int2a}
\end{equation}
\begin{equation}
\delta\omega_{E^2} = \frac{\gamma^2 E^2}{2c^{4}\omega_0} \left\langle v_x^2+v_y^2\right\rangle +O\left(1/(\omega_0 \tauc)^3\right), 
\label{eq:dw-E2-int2a}
\end{equation}
\begin{equation}
\delta\omega_{BE} = \frac{\gamma^2E}{c^2\omega_0^2}\langle b_x(0)\dot{v}_x(0) + b_y(0)\dot{v}_y(0) \rangle + O\left( 1/(\omega_0\tauc)^4 \right). 
\label{eq:dw-BE-int2}
\end{equation}
Using the expressions for the derivatives of the correlation functions
presented in the Appendix and assuming that velocities in different directions are uncorrelated and $\langle v_x^2\rangle=\langle v_y^2\rangle=\langle v_{z}^2\rangle=\frac{1}%
{3}\langle v^2\rangle$, we obtain
\begin{eqnarray}
\delta\omega_{B^2}&=&\frac{\gamma^2}{2\omega_0}\langle b_x^2+b_y^2\rangle + \frac{\gamma^2}{6 \omega_0^{3}}\langle v^2 
\rangle  \langle \vert\vec{\nabla}b_x\vert ^2 + \vert\vec{\nabla}b_y\vert ^2\rangle +O\left(  1/(\omega_0\tauc)^5 \right),
\label{eq:dw-B2-int2}
\end{eqnarray}
\begin{equation}
\delta\omega_{E^2}=\frac{\gamma^2E^2}{3c^{4}\omega_0}\langle v^2\rangle+O\left(  1/(\omega_0\tauc)^3\right),
\label{eq:dw-E2-int2}
\end{equation}
\begin{equation}
\delta\omega_{BE}=\frac{\gamma^2E}{c^2\omega_0^2}\left(\langle\frac{\partial b_x}{\partial x}\,v_x^2 \rangle + \langle\frac{\partial b_y}{\partial y}\,v_y^2 \rangle\right)  +O\left(  1/(\omega_0\tauc)^4 \right). 
\label{eq:dw-BE-int2}
\end{equation}
These results are presented in Table \ref{expressions}.
The first term in Eq. \eqref{eq:dw-B2-int2} corresponds to the leading order of the frequency shift in adiabatic regime \cite{Guigue2014}.

It is instructive to note that these results can be obtained in another way.
One can rewrite Eq. \eqref{eq:1}
\begin{equation}
S_{ij}(\omega)=\int_0^\infty e^{i\omega\tau}\langle B_{i}(0)B_{j}(\tau) \rangle \mathrm{d}\tau = \int_0^\infty e^{i\omega\tau} f(\tau) \mathrm{d}\tau \label{eq:def-Sij}
\end{equation}
and expand $f(\tau)$ in a Taylor series
\begin{eqnarray}
\nonumber
S_{ij}(\omega)  &  = & \int_0^\infty\left(  f(0) + \dot{f}(0) \tau + \cdots + f^{(n)}(0) \frac{\tau^{n}}{n!} \right)  e^{i\omega\tau}\mathrm{d}\tau \\
\nonumber
 & = & \left(  f(0) + \dot{f}(0) \frac{\partial}{\partial (i\omega) } + \cdots + \frac{f^{(n)}(0)}{n!} \frac{\partial^{n}}{\partial (i\omega)^{n}}\right)  \int_0^{\infty}e^{i\omega t}d\tau \\
 & = & f(0) \frac{i}{\omega} - \dot{f}(0) \frac{1}{\omega^2} + \ddot{f}(0)  \frac{1}{i\omega^{3}} + \cdots + f^{(n)}(0) \frac{(-1)^n}{\omega^{n+1}i^{n-1}}. 
\label{eq:Sij-dev}
\end{eqnarray}
Taking the real part (that is, the relaxation rate for the $B^2,E^2$ terms and the frequency shift for the $EB$ term, we obtain%
\begin{equation}
\mathrm{Re} \left[ S_{ij}(\omega) \right]  = -\dot{f}(0) \frac{1}{\omega^2} + \frac{\dddot{f}(0)}{\omega^{4}}+\cdots,
\label{bg1}
\end{equation}
which is equivalent to Eq. (14) of \cite{Guigue2014} and
(\ref{eq:dw-BE-int}) above but now we have another form of the correction term and
we can use this to calculate the frequency range where the first term is a good approximation.
For the imaginary part ($B^2,E^2$ frequency shifts and $EB$ relaxation) we find
\begin{equation}
\mathrm{Im} \left[ S_{ij}(\omega) \right] = \frac{f(0)}{\omega} - \frac{\ddot{f}(0)}{\omega^{3}} + \cdots ,
\label{bg2}
\end{equation}
which is equivalent to Eq. (17) of the same paper and Eq. \eqref{eq:dw-B2-int2a} above. 
In addition we find,
\begin{align}
\Gamma_{1 (B^2)} & = \gamma^2\left[  -\frac{1}{\omega_0^2}\langle b_x(0)\dot{b}_x(0)+b_y(0)\dot{b}_y(0)\rangle+\frac{1}{\omega_0^{4}}\langle b_x(0)\dddot{b}_x(0)+b_y(0)\dddot{b}_y(0)\rangle\right] , \label{bg5}\\
\Gamma_{1 (E^2)} & = \frac{\gamma^2E^2}{c^{4}}\left[-\frac{1}{\omega_0^2}\langle v_x(0)\dot{v}_x(0)+v_y(0)\dot{v}_y(0)\rangle+\frac{1}{\omega_0^{4}} \langle v_x(0)\dddot{v}_x(0)+v_y(0)\dddot{v}_y(0)\rangle\right], \\
\Gamma_{1 (BE) } & = \frac{2 \gamma^2 E}{c^2} \left[ \frac{1}{\omega_0}\langle b_x(0)v_x(\tau) + b_y(0)v_y(\tau)\rangle \mathrm{|}_{\tau=0} + \frac{1}{\omega_0^{3}}\langle b_x(0)\ddot{v}_x(0)+b_y(0)\ddot{v}_y(0)\rangle\right].
\end{align}
Using the results presented in the Appendix, we can write
\begin{align}
\Gamma_{1 (B^2)} & = -\frac{\gamma^2} {2\omega_0^2} \langle \vec{v}\cdot \vec{\nabla} \left( b_x^2 + b_y^2\right)\rangle   +O\left( 1/\left(  \omega_0\tauc\right)  ^{4}\right)  ,  \label{eq:Gamma1_B2int} \\
\Gamma_{1 (E^2)} & = -\frac{\gamma^2 E^2}{\omega_0^2c^4}\langle v_x\dot{v}_x+v_y \dot{v}_y\rangle  +O\left(  1/(\omega_0\tauc)^4 \right) \\
\Gamma_{1 (BE) } & = \frac{2\gamma^2E}{\omega_0c^2}\langle b_xv_x+b_yv_y\rangle +O\left(  1/(\omega_0\tauc)^3 \right)
\label{eq:Gamma1_BEint}
\end{align}

We expressed the correlation function derivatives in (\ref{eq:dw-B2-int2}),
(\ref{eq:dw-E2-int2}), (\ref{eq:dw-BE-int2}), \eqref{eq:Gamma1_B2int} and
\eqref{eq:Gamma1_BEint} as volume averages of the velocity and magnetic field, in the adiabatic limit. 
These expressions are therefore independent of the particle motion in the cell. 
The high frequency limits for $\delta \omega_{B^2}$, $\delta\omega_{E^2}$, $\Gamma_{1 (B^2)}$ and $\Gamma_{1 (BE)}$ are universal.

The first term in \eqref{eq:Gamma1_B2int} behaves as $1/\omega_0^2$ and  has been calculated in \cite{Guigue2014} in the diffusive adiabatic regime.
However the diffusion theory breaks down at times shorter than the collision time $\tau_{\rm{coll}}$, so that at high frequencies, ($\omega_0\tau_{\rm{coll}} \gg 1$) 
the spectrum deviates from that expected on the basis of diffusion theory. 
The high frequency (super-adiabatic) limit is correctly given by \cite{Swank2012}, using a correlation function that is valid for all times, namely the first
term in \eqref{eq:Gamma1_B2int} goes to  zero as the velocity is initially
uncorrelated with position, and the very high frequency behavior goes as
$\left(  1/\omega_0^{4}\right)$. 
The result of \cite{Swank2012} shows how the behavior goes from the $\left(  \sim1/\omega_0^2\right)$ predicted
by diffusion theory at high frequencies to $\left(  \sim1/\omega_0^4\right)$ as $\omega_0\tau_{\rm{coll}}$ becomes on the order of or greater
than 1.

\section{Nonadiabatic regime: weak magnetic field or fast particles, arbitrary fields}

We now consider the nonadiabatic limit $\omega_0 \tauc \ll 1$. 
To expand the frequency shifts expressions in terms of power of $\omega_0$,
we simply apply the same procedure of recursive integrations by parts,
changing the part that is integrated:
\begin{equation}
\int_0^\infty\sin(\omega_0\tau)f(\tau) \mathrm{d}\tau = \left[ \sin(\omega_0\tau)\int_0^\tau f(t)\mathrm{d}t\right]_0^\infty
- \omega_0 \int_0^\infty \cos(\omega_0\tau) \int_0^\tau f(t)\mathrm{d}t \mathrm{d}\tau
\label{eq:c}
\end{equation}
\begin{equation}
\int_0^\infty\cos(\omega_0\tau)f(\tau)\mathrm{d}\tau=\left[ \cos
(\omega_0\tau)\int_0^\tau f(t)\mathrm{d}t\right]_0^\infty
+\omega_0\int_0^\infty\sin(\omega_0\tau)\int_0^\tau f(t)\mathrm{d}t\mathrm{d}\tau.
\end{equation}
When applying these relations to Eq. (\ref{eq:dw-B2}), (\ref{eq:dw-E2}) and
(\ref{eq:dw-BE}), we obtain:
\begin{align}
\delta\omega_{E^2}=  &  -\frac{\gamma^2E^2}{2c^{4}}\omega_0\langle
x^2+y^2\rangle\nonumber\\
&  +\frac{\gamma^2E^2}{2c^{4}}\omega_0^2\int_0^{\infty}\sin
(\omega_0\tau)\langle x(0)x(\tau)+y(0)y(\tau)\rangle d\tau
\label{eq:dw-E2-int4}%
\end{align}%
\begin{align}
\delta\omega_{BE}=  &  -\frac{\gamma^2E}{c^2}\langle b_x x + b_y y\rangle\nonumber\\
&  +\frac{\gamma^2E}{c^2}\omega_0\int_0^{\infty}\sin(\omega_0 \tau) \langle b_x(0) x(\tau)+b_y(0) y(\tau)\rangle d\tau.
\label{eq:dw-BE-int4}%
\end{align}
One can see that the last terms on the right-hand sides of Eq. (\ref{eq:dw-E2-int4})
and (\ref{eq:dw-BE-int4}), as well as Eq. (\ref{eq:dw-B2}) involve Fourier
transforms of correlation functions which depends exclusively on position.
Since we are in the limit $\omega_0\tauc \ll 1$ we need to
consider only the first order expansion of the involved trigonometric
functions: $\sin(\omega_0\tau)\approx\omega_0\tau$. 
(For times $\tau \gtrsim \tauc$ the correlation function goes to zero.)
Applying this method to Eq. (\ref{eq:dw-B2}), (\ref{eq:dw-E2-int4}) and (\ref{eq:dw-BE-int4}), we obtain
\begin{equation}
\delta\omega_{B^2}\approx\frac{\gamma^2}{2}\omega_0\int_0^{\infty}
\tau\langle b_x(0)b_x(\tau)+b_y(0)b_y(\tau)\rangle\mathrm{d}\tau
\label{eq:dw-B2-int5}
\end{equation}%
\begin{equation}
\delta\omega_{E^2}\approx-\frac{\gamma^2E^2}{2c^{4}}\omega_0\langle
x^2+y^2\rangle+\frac{\gamma^2E^2}{2c^{4}}\omega_0^{3}\int
_0^{\infty}\tau\langle x(0)x(\tau)+y(0)y(\tau)\rangle\mathrm{d}\tau
\label{eq:dw-E2-int5}
\end{equation}
\begin{align}
\delta\omega_{BE} \approx  &  -\frac{\gamma^2E}{c^2}\langle b_x x+b_y y\rangle
\label{eq:dw-BE-int5a} \\
&  + \frac{\gamma^2E}{c^2}\omega_0^2\int_0^{\infty}\tau\langle
b_x(0)x(\tau)+b_y(0)y(\tau)\rangle\mathrm{d}\tau.
\nonumber
\end{align}

The last terms in Eq. (\ref{eq:dw-E2-int5}) and (\ref{eq:dw-BE-int5a}) can not
be calculated for any arbitrary trap geometry. But one can see that they
behave as $\omega_0^2 \tauc^2$ when $\omega_0 \tauc$ goes to zero. 
This means that the expressions of the frequency shifts are dominated by the first term on the right hand side. 
These results are presented in Table \ref{expressions}. 

Similarly, since $\cos \omega _0 \tau \approx 1$, Eq. \eqref{eq:rel-B2}, \eqref{eq:rel-E2} and \eqref{eq:rel-BE}) become
\begin{align}
\Gamma_{1 (B^2)}  & = \gamma^2\int_0^{\infty}\langle b_x(0)b_x(\tau)+b_y(0)b_y(\tau)\rangle \mathrm{d}\tau,
\label{bg1a} \\
\Gamma_{1 (E^2)}  & = \frac{\gamma^2E^2}{c^4} \omega_0^2 \int_0^\infty \langle x(0)x(\tau) + y(0)y(\tau) \rangle\mathrm{d}\tau,
\label{bg1b} \\
\Gamma_{1 (BE) }  & = \frac{2\gamma^2 E}{c^2} \omega_0 \int_0^\infty \langle b_x(0)x(\tau)+b_y(0)y(\tau) \rangle \mathrm{d}\tau,
\label{bg1c}
\end{align}
from which the low frequency limits follow immediately.

\begin{table}
\center
\begin{tabular}[c]{l|l|l}
Frequency & Adiabatic & Nonadiabatic\\
shift & (UCNs) & (Hg)\\\hline\hline
&  & \\[1pt]
$\delta\omega_{B^2}$ & $\frac{\gamma^2}{2\omega_0}\langle b_x^2+b_y^2\rangle + \frac{\gamma^2}{6\omega_0^{3}}\langle v^2\rangle  \langle \vert\vec{\nabla}b_x\vert^2 + \vert\vec{\nabla}b_y\vert^2 \rangle $ & $\frac{\gamma^2}{2}\omega_0\int_0^{\infty}\tau\langle b_x(0) b_x(\tau)+b_y(0)b_y(\tau)\rangle\mathrm{d}\tau$\\
&  & \\[1pt]
$\delta\omega_{E^2}$ & $\frac{\gamma^2E^2}{3c^{4}\omega_0}\langle v^2\rangle$ & $-\frac{\gamma^2 E^2}{2 c^{4}} \omega_0 \langle x^2+y^2\rangle$\\
&  & \\[1pt]
$\delta\omega_{BE}$ & $-\frac{\gamma^2E}{c^2\omega_0^2}\left( \langle\frac{\partial b_x}{\partial x} \, v_x^2\rangle+ \langle \frac{\partial b_y}{\partial y} \, v_y^2\rangle\right)  $ & $\frac{\gamma^2E}{c^2}\langle b_x x+b_y y\rangle$\\
\end{tabular}
\caption{Expressions of the leading terms of the frequency shifts induced by the transverse magnetic and motional fields, in the adiabatic and nonadiabatic limits.}
\label{expressions}
\end{table}

\section{\label{Sec6}Magnetic field linearly dependent on position (uniform gradients)}

In this case, which is what is usually treated theoretically and applies to most experimental situations, one can simplify the derivatives in terms of
trajectory correlation functions without any evolution equation and derive a variety of relationships. 
Let us consider a magnetic inhomogeneity $\vec{b}$ dependent linearly on the position of a spin:
\begin{align}
b_x  &  =G_x x,\\
b_y  &  =G_y y, \label{eq:uniformGradients}
\end{align}
where the relation $G_x+G_y=-\frac{\partial b_z}{\partial z} = -G_z$ 
holds by the divergence theorem. 

\subsection{General relations for fields with linear gradients and cylindrical symmetry}

\subsubsection{Expressions relating frequency shifts with frequency shifts}

In the common case of cylindrical field symmetry, (with arbitrary cell
shape) the correlation functions of interest are $\langle v_x(0)v_x%
(\tau)+v_y(0)v_y(\tau)\rangle,$ $\langle x(0)v_x(\tau)+y(0)v_y%
(\tau)\rangle,$ $\langle x(0)x(\tau)+y(0)y(\tau)\rangle$ which satisfy the
relations
\begin{align}
\langle x(0)v_x(\tau)+y(0)v_y(\tau) \rangle     & = -\frac{d}{d\tau} \langle x(0)x(\tau)+y(0)y(\tau)\rangle, \\
\langle v_x(0)v_x(\tau)+v_y(0)v_y(\tau) \rangle & = \frac{d}{d\tau}  \langle x(0)v_x(\tau)+y(0) v_y(\tau)\rangle = -\frac{d^2}{d\tau^2}\langle x(0)x(\tau)+y(0)y(\tau)\rangle.
\end{align}
Then
\begin{align}
S_{xv}\left( \omega \right) & = -i\omega S_{xx}\left( \omega \right) + \left\langle x^2+y^2\right\rangle, \\
S_{vv}\left( \omega \right) & = -i\omega S_{xv}\left( \omega \right).
\end{align}

According to Eq. (\ref{eq:dw-B2}, \ref{eq:dw-E2}, \ref{eq:dw-BE}) we find
\begin{equation}
\delta\omega_{BE} = K_{BE} \mathrm{Re} \left( S_{xv} (\omega ) \right)  = -K_{BE} \mathrm{Im} \left(  S_{vv}(\omega)/\omega\right)  = -\frac{K_{BE}}{K_{E^2}}\frac{\delta\omega_{E^2}}{\omega},
\end{equation}
with $K_{BE}=\frac{\gamma^2G_z E}{2c^2}$, $K_{B^2}=\frac{\gamma^2 G_z^2}{8}$, $K_{E^2}=\frac{\gamma^2E^2}{2c^4}$. 
The last expression was obtained in \cite{Pendlebury2004} for the case of particles moving in a cylinder with specular reflecting walls and no gas collisions, but our result holds for any cell shape and type of particle motion. 
Equation (\ref{eq:dw-BE-int4}) can be written as
\begin{align}
\delta\omega_{BE}  & = \frac{\gamma^2E G_z}{2 c^2} \langle x^2+y^2 \rangle - \frac{\gamma^2E G_z}{2 c^2}\omega_0 \int_0^{\infty}\sin(\omega_0\tau)\langle x(0)x(\tau)+y(0)y(\tau)\rangle d\tau \\
  & = \frac{\gamma^2E G_z}{2c^2}\langle x^2+y^2\rangle-\frac{4 E \omega_0}{c^2G_z}\delta\omega_{B^2}, 
\label{bg6}
\end{align}
so that Eq. (\ref{bg6}) represents a method of measuring the linear in E shift without applying an electric field. 
To do this one would apply a known constant gradient and look for a frequency shift dependent on the square of the gradient. 
The possibility of the volume average field being changed by application of the gradient can be accounted for by taking the part of the shift proportional to the square of the gradient. 
Another method would be to measure the relaxation rate due to then application of the gradient as discussed in the next paragraph. 

\subsubsection{Expressions relating frequency shifts and relaxation rates}

As the correlation functions defined in Eqs. (\ref{eq:dw-B2} - \ref{eq:rel-BE}) are all causal, that is, they are zero for $\tau<0$, their real and imaginary parts are related by a dispersion relation \cite{Papoulis} and  we can write
\begin{align}
\delta\omega_{BE}  & = K_{BE}\left[  \langle x^2+y^2\rangle-\omega \mathrm{Im}\left[  S_{xx}\left(  \omega\right)  \right]  \right] \\
& = K_{BE}\left[ \langle x^2+y^2\rangle -\frac{\omega}{\pi}\int_{-\infty}^{\infty}\frac{\mathrm{Re}\left[  S_{xx}\left(  \omega^{\prime}\right) \right]  }{\omega-\omega^{\prime}}d\omega^\prime \right] \\
& = K_{BE}\left[ \langle x^2+y^2\rangle -\frac{\omega }{\pi}\frac{1}{2 K_{B^2}}\int_{-\infty }^{\infty }\frac{\Gamma _{1\left( B^{2}\right) }}{\omega -\omega ^\prime} d\omega^\prime \right] \\
& = K_{BE}\left[ \langle x^2+y^2\rangle -\frac{1}{K_{B^2}}\frac{\omega^2}{\pi}\int_{-\infty}^{\infty}\frac{\Gamma_{1\left(  B^2\right)}\left(  \omega^{\prime}\right)  }{\omega^2-\omega^{\prime2}}d\omega^\prime \right] \label{bg3} \\
&  =K_{BE}\left[ \langle x^2+y^2\rangle -\frac{1}{K_{E^2}}\frac{\omega^2}{\pi}\int_{-\infty}^{\infty}\frac{\Gamma_{1\left(  E^2\right)}\left(  \omega^{\prime}\right)  }{\left(  \omega^2-\omega^{\prime2}\right) \omega^{\prime2}}d\omega^\prime \right]. \label{bg4}
\end{align}
Equations (\ref{bg3}) and (\ref{bg4}) are particularly interesting because they allow measurement of the frequency dependence of $\delta\omega_{BE}$, the shift, linear in $E$,
that produces a serious systematic error in the searches for particle electric
dipole moments without application of an electric field. 
By applying a gradient, $\frac{\partial b_{z}}{\partial z},$ larger than any existing gradients and measuring 
$\Gamma_{1\left(  B^2\right)  }\left(\omega\right)$ one can reconstruct the frequency dependence of 
$\delta \omega_{BE}$. 
For the case of a non-cylindrically symmetric cell we can apply
relatively large gradients $\partial b_{x,y}/\partial x,y$ and thus measure,
separately, the spectra of the correlation functions in the two directions.
While according to (\ref{bg3}) we need to know the relaxation for all
frequencies, the necessary range of measurement is limited because the known
high and low frequency limits are reached rather quickly (\ref{bg5},\ref{bg1a}). 
Substituting \eqref{eq:rel-E2} into \eqref{bg4} we obtain a form of the relation that has been obtained by another method in
\cite{Lamoreaux2005b}.

\subsubsection{Expressions relating relaxation rates with relaxation rates.}

For completeness we give relations between the relaxation rates which are abtained in a similar way: 
\begin{equation}
\Gamma_{1 (E^2) }=\frac{K_{E^2}}{K_{B^2}}\omega^2
\Gamma_{1 (B^2) }=\frac{K_{E^2}}{K_{BE}}\omega\Gamma_{1 (BE)}.
\end{equation}
The relaxation caused by the electric field alone, $\Gamma_{1 (E^2)},$ has been discussed in \cite{Schmid2008}.

\subsection{Adiabatic regime: high magnetic field or slow particles for fields
with uniform gradients}

We specify Eq. \eqref{eq:dw-B2-int2}, \eqref{eq:dw-E2-int2}, \eqref{eq:dw-BE-int2} to the case of a uniform gradient: 
\begin{equation}
\delta\omega_{B^2}=\frac{\gamma^2}{2\omega_0}\langle b_x^2+b_y^2\rangle+\frac{\gamma^2}{2\omega_0^{3}}\left\{  G_x^2\langle
v_x^2\rangle+G_y^2\langle v_y^2\rangle\right\}  +O(1/\omega_0^5\tauc^5) 
\label{eq:dw-B2-int3}
\end{equation}
\begin{equation}
\delta\omega_{E^2}=\frac{\gamma^2E^2}{2c^{4}\omega_0}\langle v_x^2+v_y^2\rangle+O(1/\omega_0^3\tauc^3)
\label{eq:dw-E2-int3}
\end{equation}
\begin{equation}
\delta\omega_{BE}=-\frac{\gamma^2E}{c^2\omega_0^2}\left\{
G_x\langle v_x^2\rangle+G_y\langle v_y^2\rangle\right\}
+O(1/\omega_0^4\tauc^4). 
\label{eq:dw-BE-int3}
\end{equation}
Similarly, the relaxation rates can be expressed as:
\begin{align}
\Gamma_{1\left(  B^2\right)  }  &  =-\gamma^2\frac{1}{\omega_0^2 }\left[  G_x^2\langle xv_x\rangle+G_y^2\langle yv_y\rangle\right]
+O(1/\omega_0^4\tauc^4) ,\\
\Gamma_{1\left(  BE\right)  }  &  =\frac{2\gamma^2E}{c^2}\left[ \frac{1}{\omega_0}\langle G_x x v_x + G_y y v_y \rangle
+O(1/\omega_0^3\tauc^3)\right].
\end{align}
The first term in Eq. \eqref{eq:dw-B2-int3} corresponds to Eq. (18) in \cite{Guigue2014}. 
It is remarkable that it is possible to derive a simple and universal expression for the third order term $\propto \omega_0^{-3}$ in Eq. \eqref{eq:dw-B2-int3}. 
To our knowledge this third order term has never been calculated before.


\subsection{Low field, high velocity limit for fields with uniform gradients}

For uniform gradients \eqref{eq:uniformGradients}, the expression of $\delta \omega_{BE}$ \eqref{eq:dw-BE-int5a} can be simplified to
\begin{align}
\delta\omega_{BE}  &  =-\frac{\gamma^2E}{c^2}\langle G_x x^2+G_y y^2\rangle+\frac{\gamma^2E}{c^2}\omega_0^2\int_0^{\infty}\tau\langle G_xx(0)x(\tau)+G_yy(0)y(\tau)\rangle\mathrm{d}\tau\label{eq:dw-BE-int-unif}\\
&  =-\frac{\gamma^2E}{c^2}\langle G_xx^2+G_yy^2\rangle+O(\omega_0^2\tauc^2) .
\end{align}
Also \eqref{eq:dw-B2-int5} becomes
\begin{equation}
\delta\omega_{B^2}\approx\frac{\gamma^2}{2}\omega_0\int_0^{\infty}\tau\langle G_x^2x(0)x(\tau)+G_y^2y(0)y(\tau)\rangle\mathrm{d}\tau
\end{equation}
and \eqref{eq:dw-E2-int5}
\begin{equation}
\delta\omega_{E^2}=-\frac{\gamma^2E^2}{2c^{4}}\omega_0\langle x^2+y^2\rangle+\frac{\gamma^2E^2}{2c^{4}}\omega_0^{3}\int_0^{\infty}\tau\langle x(0)x(\tau)+y(0)y(\tau)\rangle\mathrm{d}\tau .
\end{equation}
Similarly
\begin{align}
\Gamma_{1 (B^2)} & = \gamma^2 \int_0^{\infty}\langle G_x^2x(0)x(\tau)+G_y^2y(0)y(\tau)\rangle\mathrm{d}\tau ,\\
\Gamma_{1 (E^2)} & = \frac{\gamma^2E^2}{c^4}\omega_0^2\int_0^{\infty}\langle x(0)x(\tau)+y(0)y(\tau)\rangle\mathrm{d}\tau ,\\
\Gamma_{1 (BE) } & = \frac{2 \gamma^2E}{c^2}\omega_0\int_0^{\infty}\langle G_xx(0)x(\tau)+G_yy(0)y(\tau)\rangle\mathrm{d}\tau, 
\end{align}
so that in the low frequency limit there are no universal or quasi-universal expressions.

\section{Conclusion}

In this paper we have investigated the  asymptotic behavior of the spin-relaxation and related frequency shifts due to the restricted motion of  particles in non-uniform magnetic and electric fields. 
Simple universal expressions (valid for any form of gas container and any spatial form of the field) were obtained for the observables $\delta\omega$ and $\Gamma _1$ for adiabatic and nonadiabatic regimes of spin - motion. 
The remarkable feature of all our results is that they were obtained without any specific assumptions about the explicit form of the correlation functions. 
Hence, we expect that our results are valid for both diffusive and ballistic regimes of motion. 
These results can then be applied to a wide variety of realistic systems. 
They are especially important in the context of experiments searching for the electric dipole moment using trapped particles, for the frequency shifts proportional to electric fields are of utmost importance. 
In particular we have given general relations between various frequency shifts and relaxation rates. 

\section{Acknowledgment}
The authors would like to thank Albert Steyerl for pointing out a numerical error in the first verision of the manuscript. 

\appendix

\section{\label{app:corrfuncder} Useful expressions of correlation functions}

We derive three useful relations for the derivative of certain correlation functions in terms of volume averages. 
Let us consider an inhomogeneity $b_i$ with $i=x$ or $y$. 
\begin{align}
\der{}{\tau}\langle b_i(0)b_i(\tau )\rangle \vert_{\tau=0} &=\langle b_i(0)\dot{b}_i(0)\rangle \\  
&= \langle b_i\vec{\nabla}b_i \cdot \der{\vec{r}}{\tau}\rangle \\
&  = \langle b_i\frac{\partial b_i}{\partial x}v_x\rangle+ \langle
b_i\frac{\partial b_i}{\partial y}v_y\rangle+ \langle b_i\frac{\partial
b_i}{\partial z}v_{z}\rangle\label{eq:bbdot}\\
& = \frac{1}{2} \langle \vec{v}\cdot \vec{\nabla} b_i^2\rangle%
\end{align}
\begin{align}
\derr{}{\tau}\langle b_i(0)b_i(\tau )\rangle \vert_{\tau=
0} &=\langle b_i(0)\ddot{b_i}(0)\rangle \\
&  = -\langle\dot{b_i}(0)\dot
{b_i}(0)\rangle\\
&  = -\langle\left(  \vec{\nabla}b_i\cdot\frac{\mathrm{d}\vec{r}}{\mathrm{d}%
\tau}\right)  ^2\rangle\\
&  = \langle\left(  \frac{\partial b_i}{\partial x}v_x \right)^2 \rangle+
\langle\left(  \frac{\partial b_i}{\partial y}v_y \right)^2 \rangle+
\langle\left(  \frac{\partial b_i}{\partial z}v_{z}\right)^2 \rangle\\
&  = \langle\left(  \frac{\partial b_i}{\partial x}\right)^2 \rangle\langle
v_x^2\rangle+ \langle\left(  \frac{\partial b_i}{\partial y}\right)^2\rangle\langle v_y^2\rangle+ \langle\left(  \frac{\partial
b_i}{\partial z}\right)^2\rangle\langle v_z^2\rangle\label{eq:bdotbdot}
\end{align}
\begin{align}
\frac{\mathrm{d}}{\mathrm{d}\tau}\langle b_i(0)v_{j}(\tau)\rangle\vert_{\tau=0}  &  = - \langle\dot{b_i(0)}v_{j}(0)\rangle= -\langle(\vec{\nabla}b_i\cdot
\vec{v})v_{j}\rangle\\
&  = - \langle\rder{b_i}{j} \, v_{j}^2\rangle\label{eq:bdotv}%
\end{align}

\bibliographystyle{apsrev}
\bibliography{article}
\end{document}